\documentclass[atmosphere,article,accept,pdftex,moreauthors]{Definitions/mdpi} 
\usepackage{amsthm,amsmath,amssymb}
\usepackage{mathrsfs}
\firstpage{1} 
\pubvolume{1}
\issuenum{1}
\articlenumber{0}
\pubyear{2023}
\copyrightyear{2023}
\externaleditor{Academic Editor: Firstname Lastname}
\datereceived{10 August 2023} 
\daterevised{13 September 2023} 
\dateaccepted{17 September 2023} 
\datepublished{ } 
\hreflink{https://doi.org/} 

\Title{On the Impact of ENSO Cycles and Climate Change on Telescope Sites in Northern Chile}

\TitleCitation{On the Impact of ENSO Cycles and Climate Change on Telescope Sites in Northern Chile}

\Author{Julia Victoria Seidel $^{1,}$*$^{,\dagger,}$\orcidA{}, Angel Otarola $^{1}$ and  Valentina Th\'eron $^{1,2}$}

\AuthorNames{Julia Victoria Seidel and Angel Otarola and Valentina Th\'eron}

\AuthorCitation{Seidel, J.V.; Otarola, A.; Th\'eron, V.}

\externaleditor{Academic Editor: Marc Sarazin}

\address{%
$^{1}$ \quad European Southern Observatory, Alonso de C\'ordova 3107, 763000, Santiago de Chile, Chile; angel.otarola@eso.org (A.O.); valentina.theron@free.fr (V.T.)\\
$^{2}$ \quad Ecole Nationale de la Météorologie, 42 Avenue Gaspard Coriolis, BP 45712 31057 Toulouse, France}

\corres{Correspondence: jseidel@eso.org}

\firstnote{ESO Fellow.} 

\abstract{The Atacama desert stands as the most arid, non-polar, region on Earth and has accommodated a considerable portion of the world's ground-based astronomical observatories for an extended period. The comprehension of factors important for observational conditions in this region, and the potential alterations induced by the escalating impact of climate change, are, therefore, of the utmost significance. In this study, we conduct an analysis of the surface-level air temperature, water vapour density, and astronomical seeing at the European Southern Observatory (commonly known by its acronym, ESO) telescope sites in northern Chile. Our findings reveal a discernible rise in temperature across all sites during the last decade. Moreover, we establish a correlation between the air temperature and water vapour density with the El Ni\~no Southern Oscillation (ENSO) phases, wherein, the warm anomaly known as El Ni\~no (EN) corresponds to drier observing conditions, coupled with higher maximum daily temperatures favouring more challenging near-infrared observations. 
The outcomes of this investigation have potential implications for the enhancement of the long-term scheduling of observations at telescope sites in northern Chile, thereby aiding in better planning and allocation of resources for the astronomy community.}


\begin{document}



\section{Introduction}
\label{sec:intro}
The Atacama climate is primarily influenced by large-scale subsidence at subtropical latitudes. However, the~combination of the Andean slope and the adjacent cold Humboldt stream in the Pacific Ocean also makes a significant contribution~\cite{Rutllant2003}. This interaction results in an atmospheric marine boundary layer characterised by extremely dry and cloud-free air above it (relative humidity below $15\%$), providing favourable conditions for astronomical observations \citep{Munoz2011}. 

It is precisely this fortuitous geography, with~extremely dry and cloud-free air at high altitudes that makes the Atacama desert the primary host of world-class astronomical facilities in the world. Not only does northern Chile already host the vast majority of 8~m class visible and near-infrared telescopes, but~it is also the construction site of the world's first 40~m class telescope, the~Extremely Large Telescope (ELT).
Important factors for the data quality in cutting-edge astronomy are the integrated absolute humidity (precipitable water vapour, PWV), together with meteorological conditions, such as the surface air temperature, wind speed, and~relative humidity \citep{TMT2009, ELT2011}. PWV and its proxy, the surface water vapour density (WVD), are relevant in the near- to mid-infrared range, as~water vapour affects the astronomical observations via atmospheric absorption and thermal emissions \citep{Bustos2014, Otarola2015}. This means the absorption of light by the water in Earth's atmosphere significantly reduces the light received by telescopes, especially in the near-infrared. Observing schedules at the European Southern Observatory, the~focal point of this work, are created up to half a year in advance when weather conditions are still unclear. Therefore, if~too many observational programmes with strong constraints on precipitable water vapour are scheduled, but~then atmospheric conditions do not allow for them, the~telescopes will end up with fewer observable programmes in the scheduling period than were needed. In~other words, during~observing nights with adequate but not excellent observing conditions, where less restrictive observations could have been observed, the~telescopes are needlessly sitting idle. This lost time is aptly called idle time and should be avoided as much as possible. To~this date, the~astronomy community has no metric on which to base the average conditions beyond the known annual seasonal variations, and~the impact of climate change on the future of infrared programmes is uncertain at best. 

\noindent In addition to the scheduling issue, the~overall quality of observations during a specific night is an important metric to assess the suitability of an observing site. In~the visible and near-infrared, astronomers employ the \textit{seeing} term for this purpose. In simple terms, seeing is a measure of the angular broadening of an astronomical point-like source on the image plane due to Earth's atmospheric turbulence. It is expressed in \textit{arcseconds}, a unit of small angles common in astronomy, where one arcsecond is equal to $\frac{1}{3600}$ of a sexagesimal degree. The~smaller the seeing magnitude the~better resolved the image of the observed object. Excellent seeing for astronomical research is defined as seeing below $0.6"$ and indicates a near-motionless image. Median level seeings range from $0.6"$ to $0.9"$. Acceptable seeing ranges from $1.5"$ up to $2.0"$ at which levels the image starts to become more distorted. Important factors affecting the turbulent mixing of the atmospheric layers, and hence affecting the seeing, are the vertical gradients of the horizontal wind speed and temperature gradients \citep[][and references therein]{Melnick2009}. It remains a topic of much interest to know whether astronomical seeing is affected by long-term climate oscillations or climate change, a point to which we aim to contribute in this manuscript. In~the context of the long timescale and associated cost for the construction and the subsequent operation of the ELT in the next half century, an~assessment and hopefully a predictive tool of overall atmospheric conditions and seeing is crucial for the community.

In this work, we aim to fill this knowledge gap for the astronomy community by assessing the suitability of ENSO as a predictive tool for interannual condition changes. The prominent feature of the El Ni\~no (EN) phase of ENSO is the warming of the tropical Pacific, quantified by the Ni\~no3.4 index, leading to increased near-surface air temperatures down to 30$^\circ$ S—encompassing all observatories in this study along the Pacific coast of South America. Its counterpart, La Ni\~na (LN), is marked by opposite conditions. While EN phases have been correlated with higher precipitation in the  30$^\circ$--41$^\circ$ S latitudes band~\citep{Montecinos2003}, the~subtropical region, where the Atacama Desert lies, has been correlated with dryer conditions~\citep{RojasMurillo2022}.

Hence, we assess the impact of ENSO phases and global warming trends on WVD and temperature for European Southern Observatory sites, providing essential metrics for the community to evaluate long-term observational conditions in northern Chile. We propose strong ENSO phases as a metric to better schedule challenging observing programmes, but~find no significant impact of global warming on astronomical site quality. However, we confirm the globally seen warming trend and explore local effects on high altitude sites and curious local cooling effects impacting the Paranal Observatory specifically. With regards to seeing, we show that the median seeing conditions remain stable within the natural inter-annual variability, a crucial result in the context of the ELT.
The manuscript is structured as follows: In Section~\ref{sec:data}, we provide an overview of the sites, the~acquired data, and~the analysis methods. In~Section~\ref{sec:results}, we link the data to climate indices and explore the connection to climate change. We then provide important conclusions for the astronomy community in Section~\ref{sec:concl} where we highlight the power of established climate indices to prevent time loss due to sub-optimal planning and provide an outlook for future studies to monitor seeing in a warming~climate.

\section{Materials and~Methods}
\unskip
\label{sec:data}
\subsection{Data Collection and~Accessibility}

For this study, we collected data from three different observatories at four different sites (see overview in Table~\ref{tab:sites}). We provide instructions on how to access all the data in this study in the Data Availability Statement at the end of the manuscript. An overview of the coverage and further details are presented in Table~\ref{tab:overviewData}.

\begin{table}[H] 
\caption{Location of all European Southern Observatory sites used for data collection in this~study.}\label{tab:sites}
\begin{tabular*}{\textwidth}{lccc}
\toprule
\textbf{Observatory}	& \textbf{Latitude [\boldmath$^\circ$S]}	& \textbf{Longitude [\boldmath$^\circ$W]} & \textbf{Elevation [MASL]}\\
\midrule
la Silla	& 29.258	& 70.738 & 2400 \\
Paranal		& 24.627 & 70.404 & 2650 \\
Llano del Chajnantor (CBI) & 23.0333 & 67.7667 & 5080 \\
Llano del Chajnantor (APEX) & 23.0058 & 67.7592 & 5150 \\
\bottomrule
\end{tabular*}
\end{table}
\unskip

\begin{table}[H]
\caption{Data provided from the respective meteo station (T and RH at 2m)\label{tab:overviewData}}
\setlength{\cellWidtha}{\columnwidth/5-2\tabcolsep-0.4in}
\setlength{\cellWidthb}{\columnwidth/5-2\tabcolsep+0.4in}
\setlength{\cellWidthc}{\columnwidth/5-2\tabcolsep+0.1in}
\setlength{\cellWidthd}{\columnwidth/5-2\tabcolsep-0in}
\setlength{\cellWidthe}{\columnwidth/5-2\tabcolsep-0.1in}
\scalebox{1}[1]{\begin{tabularx}{\columnwidth}{>{\PreserveBackslash\centering}m{\cellWidtha}>{\PreserveBackslash\centering}m{\cellWidthb}>{\PreserveBackslash\centering}m{\cellWidthc}>{\PreserveBackslash\centering}m{\cellWidthd}>{\PreserveBackslash\centering}m{\cellWidthe}}
\toprule

			\textbf{Site}		& \textbf{Obs Period}	& \textbf{Time Resolution}  & \textbf{$\#$Days with Obs}  & \textbf{\% of Coverage}   \\
			\midrule
                la Silla	& 10-May to 23-April			& 1 min        &   4740  & $91.20\%$\\
			  	                
                    \midrule
\multirow{2}{*}{Paranal}			& 98-July to 23-April	& 1 min    & 13,997    \textsuperscript{1}   & $91.86\%$ \textsuperscript{1}\\
      	& 84-November to 98-June & 20 min to h +    & -    & -\\
                   \midrule
CBI		& 99-April to 5-July		& 1 h      & $2867$    & $77.01\%$\\

APEX		& 6-January to 23-April			& 1 min    & $6050$    & $98.61\%$\\
			  	                 
			\bottomrule
		\end{tabularx}}
  \noindent{\footnotesize{\textsuperscript{1} Combining historical and current data.}}
\end{table}

\subsubsection{La~Silla}

The Astronomical Weather Station (AWS) started routine operation in September 1991 and full automation was reached in December 1993. From~October 1988 to February 1991, measurements were conducted from Cerro Vizcachas, 7~km to the east of La Silla, which was once a candidate site for the Very Large Telescope (VLT). In~2008, the~AWS was struck by lightning and had to be replaced by a new weather station and sensors. This replacement was completed in May 2010. However, the~data before this event show a marked offset in pressure, jumping from previously approximately 774~hPa median pressure to 769~hPa median pressure. This change is also visible in different temperatures and dew points before and after the event. It is unclear whether this difference is due to a change in the location of the AWS (no such change was recorded) or if the values were previously recorded at 30~m and then at 2~m, but~saved in the database as a continuous series at 2~m. Due to this inconsistency, we focus our study on the data after the lightning strike, with~the new AWS station (2010 and beyond).

\subsubsection{Paranal~Observatory}
The temperature at different altitudes for the Paranal Observatory, as~well as the relative humidity and the wind direction, were obtained from the European Southern Observatory ambient query form. However, the~ambient query form only contains data from the Vaisala Meteorological station at the Paranal Observatory since August 1998 when the station was upgraded. We obtained the historical values since its installation in October 1984 from European Southern Observatory internal documents. Measurements were suspended from July 1991 to September 1992 for the 14-month duration of the levelling work of the Paranal Observatory peak to make the platform needed to install the VLT. No offsets were found in the data due to this intervention. Values are sparse in the historical data until April 1994 (one or two daily measurements) and then show a 20-minute cadence. The~sparse data are not taken into account in our analysis but are available upon request. No change points were found between the historical data and the current~data.

The optical turbulence at the Paranal Observatory is measured with the help of two instruments that share the same optical tube assembly. The~integrated turbulence along the atmospheric column is measured with a differential image motion monitor (DIMM) instrument~\citep{Sarazin1990}. Sharing the field-of-view is the multi-aperture scintillation sensor (MASS) instrument \citep{Tokovinin2007}, which provides a measure of the vertical profile of the optical turbulence. The~MASS helps to obtain the integrated seeing in the troposphere, 250m above the surface level and beyond. Hence, these instruments have proven important to monitor both the~integrated turbulence in the whole atmosphere above the DIMM level (currently 7~m above the surface) and the seeing in the~free-troposphere.

\subsection{Paranal Observatory Precipitable Water Vapour and Water Vapour Density~Calculation}

\begin{figure}[H]
\includegraphics[width=11.8cm]{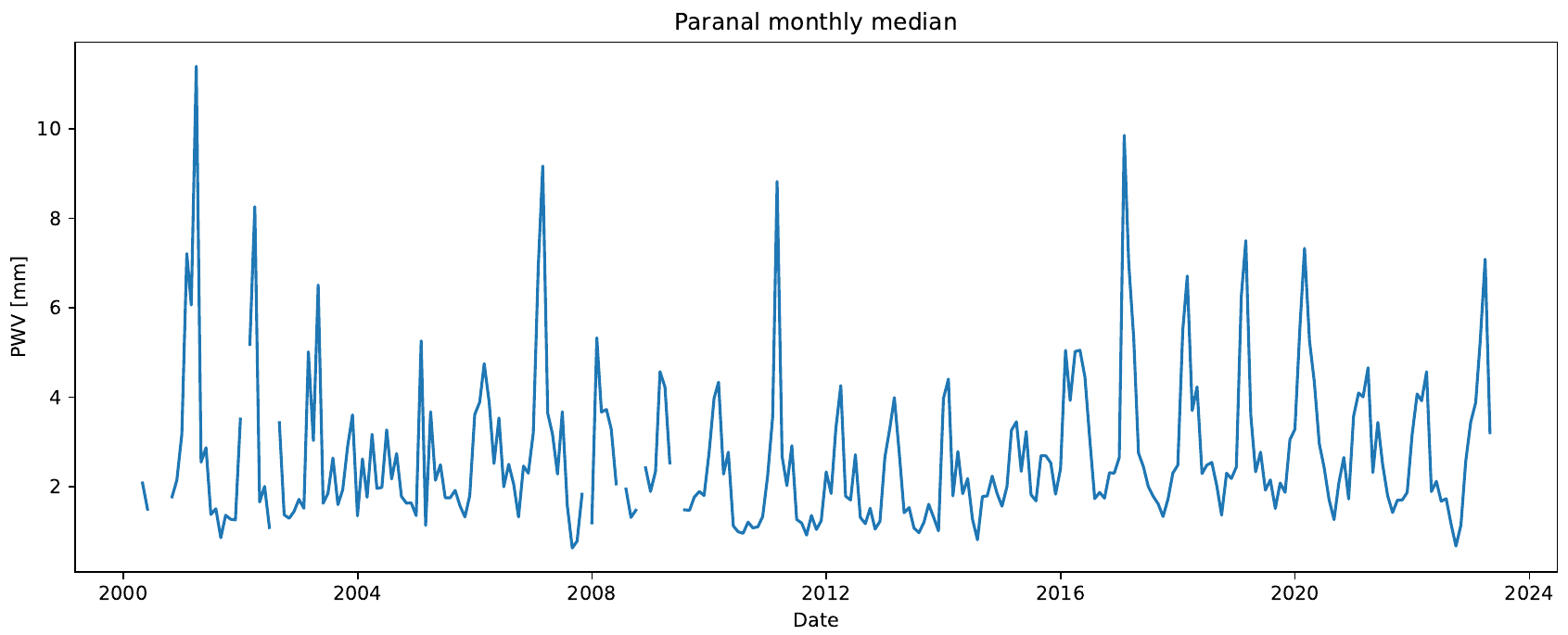}
\caption{Monthly median PWV at the Paranal Observatory. No significant incremental trends for minimum PWV are discernible. However, for~the annual minimum PWV, a marginal seasonal pattern on a decadal scale might be inferred. This pattern remains statistically~non-significant.} \label{fig:PWVParanal}
\end{figure}

The PWV measurements from the Paranal Observatory were obtained via two sources. Starting from April 2015, PWV measurements were performed with the LHATPRO (Low
Humidity And Temperature PROfiling, \citep{Kerber2012}) instrument. This is a microwave radiometer covering the strong water vapour emission line at 183~GHz, and~providing a record of measurements with a time resolution of a few seconds. Historical values are available at a lower time sampling from the observation of the equivalent width telluric standard stars under clear sky conditions with the XSHOOTER and UVES instruments. PWV measurements from XSHOOTER, at~an average sampling rate of every three hours, are available from October-2009 to March-2020, while one daily measurement or less from UVES is available from April-2000 to September-2020. The~data, as monthly median values for the Paranal Observatory, are shown in Figure~\ref{fig:PWVParanal}; no augmentation in minimum PWV with time is discernible. When data from more than one measurement were available, we verified that the difference between the data was less than $1\sigma$ and then provided the mean in the plot. No conflicting data points were found and the measurements from XSHOOTER and UVES confirmed the new LHATPRO measurements. The~PWV of the Chajnantor site has been studied in detail in~\cite{Cortes2020}. Unfortunately, for~the la Silla Observatory, no such observations exist. Given that we are mainly interested in the impact on surface conditions at telescope sites, we use the water vapour density (WVD) as a proxy for all three~sites.

The water vapour density (WVD) is calculated from the relative humidity (RH) and the temperature (T) at the altitude of the meteorological sensors station. The~water vapour pressure at saturation $e_\mathrm{sat}$ is derived from these two parameters via the solution of the Clausius--Clapeyron equation from~\cite{Buck1981}:
\begin{equation}
    e_\mathrm{sat} = 6.1121 \cdot \exp{\frac{17.502 T_c}{240.97+T_c}} \cdot 100 \mathrm{[Pa]}
\end{equation}
where $T_c$ is the temperature in degrees Celsius. The~nominal water vapour pressure is then
\begin{equation}
    e_0 = e_\mathrm{sat} \cdot \frac{\mathrm{RH}}{100} \mathrm{[Pa]}
\end{equation}
and, following the ideal gas law,
\begin{equation}
    \mathrm{WVD} = \frac{e_0}{R_v T_K} \cdot 1000 [\frac{\text{g}}{\text{m}^3}]
\end{equation}
with $T_K$ the temperature in Kelvin, and~$R_v$ as the specific gas constant for water vapour ($461.5~\frac{J}{K \cdot \mathrm{kg}}$), the~WVD can be derived for all three observatory~sites.

\subsection{Paranal Observatory Turbulence~Quantification}

We use the computed surface level temperature and wind speed gradients to produce an overall estimate of the Richardson number as a proxy for turbulence. The~Richardson number is the result of taking the ratio between the buoyancy forces and the square of the wind shear, and~is computed as follows:
\begin{equation}
    \mathscr{R}i = \frac{g}{\theta} \frac{\frac{d\theta}{dh}}{(\frac{dU}{dh})^{2}} \mathrm{[adimentional]}
    \label{eq:richardson}
\end{equation}
\begin{equation}
    \theta = T\mathrm{[K]}(\frac{P_{0}}{P})^{R/c_{p}} \mathrm{[K]}
\end{equation}
where $g$ is the acceleration of gravity (9.8~$\frac{\text{m}}{\text{s}^2}$),  $\theta$ is the mean air potential temperature, $d\theta/dh$ is the vertical gradient of air potential temperature, and $dU/dh$ is the vertical gradient of the horizontal wind speed. $P_{0}$ is the pressure at a reference altitude level (the sea level, $\sim$1013~hPa), R is the gas constant of mixed air ($287.05$~$\frac{\text{J}}{\text{K} \cdot \mathrm{kg}}$), and $c_{p}$ is the specific heat capacity at constant pressure ($1005.0$~$\frac{\text{J}}{\text{K} \cdot \mathrm{kg}}$).  

\subsection{Missing Values and~Consistency}

Some of the historical data show sections of missing data, some as large as years, but~mostly one value or two during the day. In~the cases where only one value was missing consecutively, we applied a linear interpolation before building monthly sums or means. In~cases where more than one value was missing in a row, the~missing value was treated like a propagating NaN. 
For consistency, we checked that the maximum daily value was not below the minimum daily value of the previous day \citep{Hunziker2017}. No inconsistent values were found. In~order to avoid an undue influence of outliers on our analysis of the minimum and maximum values, we calculate extremes as the $0.5\%$ and $99.5\%$ percentile values,~respectively.

\subsection{Trends and Correlation~Calculation}

The seasonal trends were calculated for the austral seasons: summer (December to February, DJF), autumn (March to May, MAM), winter (June to August, JJA), and~spring (September to November, SON). The~summer season thus commences in the December of the previous calendar year. As~we are principally interested in the impact on astronomical sites, the~data were separated into day and night by calculating the time when the sun zenith angle was $90^\circ$ at the telescope site (i.e., above the horizon). 
Considering that temperatures usually follow non-normal distributions, we applied the non-parametric Mann--Kendall test to estimate trends \citep{Kendall1949} with the removal of potential auto-correlation in time series \citep{Yue2004}, as implemented in \citep{Hussain2019}. In~this implementation, trends are established with a Kendall's $\tau$-based slope estimator \citep{Sen1968}, with~statistical significance set to the $5\%$ level (\emph{p}-value).
We estimated the correlation between our observed parameters and modes of climate variability with a Student's \emph{t}-test. The~resulting Pearson's correlation coefficient (PCC) is shown, together with the \emph{p}-value. The~significance level of the correlation is set at $5\%$ (\emph{p}-value).

\subsection{Climate~Indices}

The general atmospheric circulation can be disturbed by internal modes---large-scale disruptions due to the unstable coupling of the atmosphere--ocean system. This cyclical climate variability plays an important role in South America's climate on both sides of the Andean Cordillera, as~recently highlighted in~\cite{Gomez2023}, where changes in the surface wind of western Patagonia were linked to ENSO and PDO modes. The~three most important large-scale modes of climate variability in South America are described as the Antarctic circulation ({AAO,}  \citep{Thompson2000}), and~the aforementioned long-term Pacific decadal oscillation ({PDO,}~\citep{Mantua1997}) and El Niño-Southern Oscillation ({ENSO,}~\citep{Smith2000}). In~this work, we will mostly focus on ENSO and provide only a surface definition of the other two phenomena and their~impact.

The AAO is the main component of the 850~hPa geopotential height anomalies south of $20^\circ$S but also influences temperature and precipitation anomalies in South America, e.g., the~AAO is associated with precipitation anomalies in southern Chile and the subtropical east coast of the continent \citep{Garreaud2009}. The PDO is the main reason for the monthly sea surface temperature anomalies in the Pacific Ocean (north of 20$^\circ$ N) with a decadal oscillation between warm and cold phases. The ENSO operates on shorter cycles ranging between 2 and 7~years and is typically separated into two anomaly cycles: El Niño (EN, warm phase) and La Niña (LN, cold phase), where occurrences in the EN phases are usually roughly inverse to the LN phases. The ENSO is monitored via various indices with the Niño3.4 index the most appropriate choice for northern Chile \citep{Boehm2020}. Niño3.4 quantifies the SST anomaly between 5$^\circ$ N--5$^\circ$ S and between 170--120$^\circ$ W and was obtained from the National Oceanic and Atmospheric Administration of the U.S. Department of Commerce as a monthly mean time series. An~EN or LN phase is declared when the temperature anomaly deviates $>|0.5\,^\circ\text{C}|$ from the historical mean. We define strong LN or EN phases by a deviation of $>|0.75\,^\circ\text{C}|$ from the historical~mean.

\section{Results}
\label{sec:results}
\unskip

\subsection{European Southern Observatory Sites Temperature~Trends}

All sites in our study are found at geographical elevations 2400 m above sea level (MASL). However, both the la Silla and the Paranal Observatory are close to the coastal line and at the lower boundary of high-elevation sites. On~the other hand, the~Chajnantor site lies at over 5000~MASL on the western slope of the Andes mountains. This allows us to provide some insights into how temperature trends change with altitude. The~significance of temperature trends, as~a function of site (altitude), was assessed using the Mann--Kendall (MK) test. See Appendix~\ref{app:MKtest} for the tables for all sites together with a more in-depth analysis of the temporal warming variations for the  Paranal Observatory (Appendix~\ref{app:MKtest}-Tables \ref{tab:ParanalT} and \ref{tab:ParanalTcurrent}).

Our results show a significant increase in temperatures at the Chajnantor site, with~rising trends for temperatures during the austral summer. These trends confirm the projected altitude-dependent warming acceleration, both globally \citep{Pepin2016} and for future observatories~\cite{Haslebacher2022}. In~the austral autumn, the~minimum temperatures at the Chajnantor site exhibit an increase of $~1\,^\circ\text{C}/$decade, twice as fast as any other parameter. A~consistent rise in the minimum temperature during the austral autumn (and the maximum temperature in summer) extends across all sites. Previous studies on climate trends in Chile have demonstrated substantial warming during the austral autumn and summer while suggesting cooling during spring and winter \citep{Burger2018} linked to climate change \citep{Rangwala2012}. Our findings corroborate a strong warming trend, specifically evident in the minimum temperature during autumn, which implies reduced cool periods leading up to winter. This indicates a lengthening of the ablation season in the high Andes with an impact on the glacier mass balance \citep{Burger2018}. Nonetheless, the climate forcing responsible for glacial shrinkage in Chile is a complex subject \citep{Masiokas2016} and beyond the scope of this~work. 

\begin{figure}[H]

\begin{adjustwidth}{-\extralength}{0cm}
\centering 
\includegraphics[width=14.8cm]{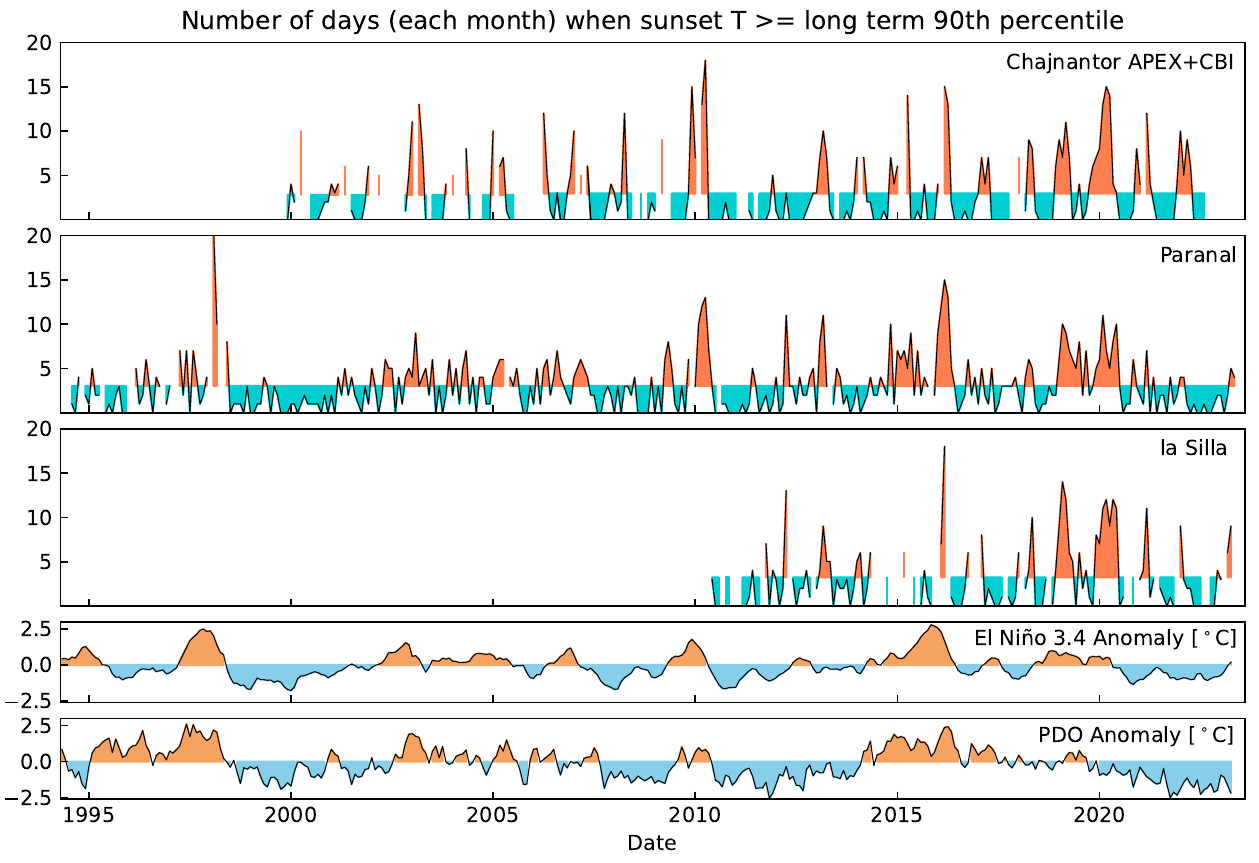}
\end{adjustwidth}
\caption{Days in a month with the ambient temperature at 2m above surface above the long-term 90th percentile for all sites studied. Values below the thresholds are shaded in blue, and~values above the threshold are shaded in red. The~lowest two panels show the Niño3.4 index and the PDO (Pacific Decadal Oscillation) index. \label{fig:TaENSO}}
\end{figure}

While a definitive warming trend attributed to global warming is evident across European Southern Observatory sites, annual-level variations cannot be solely ascribed to climate change. We investigated the link between extreme warm events when the daily maximum ambient temperature exceeded the 90th percentile of the long-term temperature record and ENSO (Figure \ref{fig:TaENSO}). Our findings reveal a distinct correlation between the occurrence of warm surface temperature anomalies and the Niño3.4 index across sites. Further details on the correlations for all sites can be found in Table \ref{tab:TaENSOCorr} and in Appendix~\ref{app:MKtest}.

\unskip  

\begin{table}[H] 
\caption{Student's \emph{t}-test for correlation between El Niño3.4 anomaly and sunset temperature outliers (see Figure~\ref{fig:TaENSO}). \label{tab:TaENSOCorr}}
\newcolumntype{C}{>{\centering\arraybackslash}X}
\begin{tabularx}{\textwidth}{CCCC}
\toprule
\textbf{Observatory}	& \textbf{PCC}	& \textbf{\emph{p}-Value} & \\
\midrule
Chajnantor (CBI+APEX)		& $0.192$			& $0.004$ & correlation\\
Paranal		& $0.453$			& $0.000$ &  correlation\\
la Silla		& $0.228$			& $0.012$ & correlation\\

\bottomrule
\end{tabularx}
\end{table}
\unskip

\subsection{European Southern Observatory Sites Surface Water Vapour Density Trends and Precipitable Water~Vapour}

\begin{figure}[H]
\includegraphics[width=13.8cm]{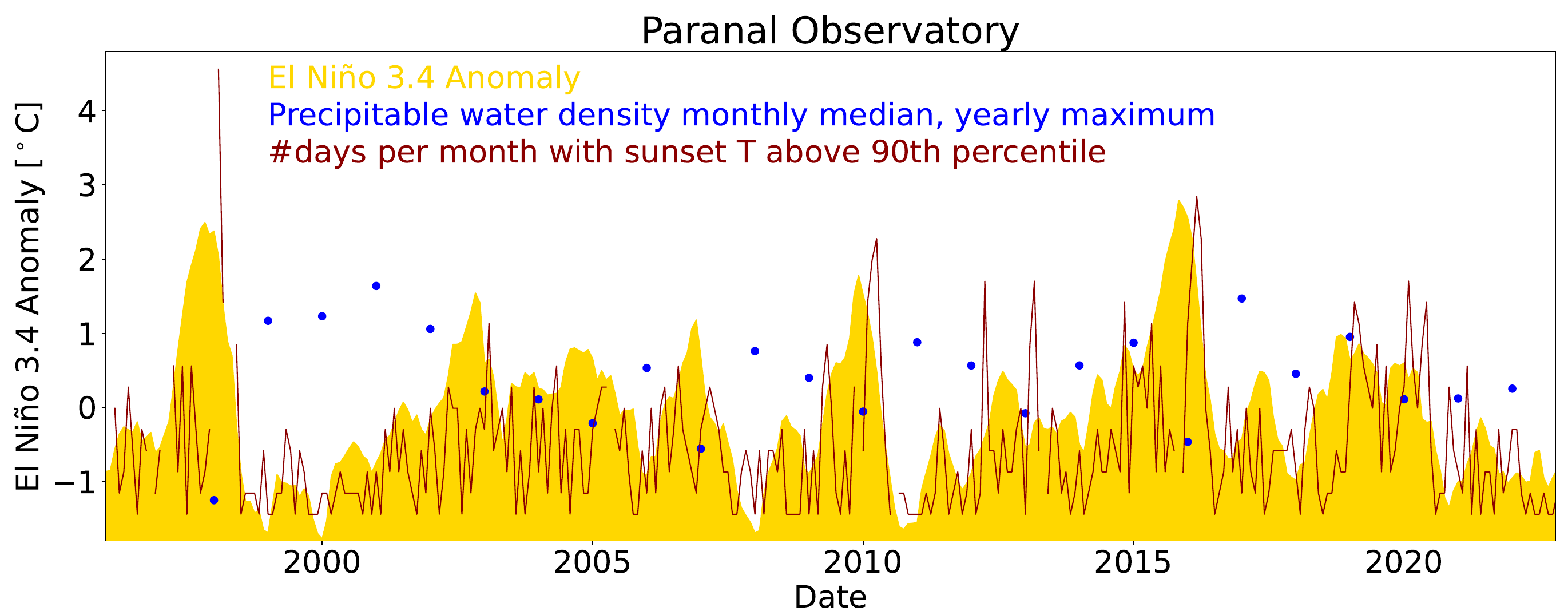}
\caption{The El Niño3.4 anomaly is shown in yellow with the number of days with sunset temperature above the 90th quantile (red line) and the WVD (blue dot) overlaid on the same scale as EN3.4 for the Paranal Observatory. The~WVD is displayed as the wettest yearly period, the~month each year with the highest mean WVD. A~clear correlation is visible between positive ENSO anomalies and extremely warm and dry~periods. \label{fig:ParanalTrends}}
\end{figure}

The surface water vapour density (WVD) in the Paranal and the la Silla Observatories is anti-correlated with ENSO (see Table \ref{tab:WVDENSOCorr}). Years during strong EN phases show no surface-level wet periods, while there are a larger number of days with temperatures above the long-term 90th quantile, e.g.,~in 1998, in~2010, and in 2016 (see Figure~\ref{fig:ParanalTrends}). While a decrease in WVD is also observed for the Chajnantor site during strong EN phases, the~overall WVD is extremely low, and~no correlation could be determined in our short data series. The~most important site to quantify is the Paranal Observatory due to its close proximity to the construction site (Cerro Armazones) of the future Extremely Large Telescope (ELT)~\cite{Gilmozzi2007}. 

The monthly median values of precipitable water vapour (PWV) for the Paranal Observatory are shown in Figure~\ref{fig:PWVParanal} and no augmentation in minimum PWV with time is discernible; however, a~marginal decadal oscillation can be inferred. If~anything, what is clearly visible is the variability in the maximum PWV events during the austral summers. This variability is most likely influenced by the predominant position of the Bolivian high-pressure centre during summer months~\cite{Garreaud2011, Krishnamurti2013, Segura2020}, which leads to dry anomalies during the normally wet summer for the combination of EN/PDO(+) \cite{RojasMurillo2022}. The~PWV long-term record for the Chajnantor site was analysed in~\cite{Cortes2020}, and~no evidence of PWV trends over the 20 years of data was found, in~agreement with our results for the Paranal~Observatory.

\unskip

\begin{table}[H] 
\caption{Student's \emph{t}-test for correlation between el Niño3.4 anomaly annual maximum value of WVD monthly~medians.} \label{tab:WVDENSOCorr}
\setlength{\cellWidtha}{\columnwidth/4-2\tabcolsep-0.0in}
\setlength{\cellWidthb}{\columnwidth/4-2\tabcolsep-0.0in}
\setlength{\cellWidthc}{\columnwidth/4-2\tabcolsep-0.0in}
\setlength{\cellWidthd}{\columnwidth/4-2\tabcolsep-0.0in}
\scalebox{1}[1]{\begin{tabularx}{\columnwidth}{>{\PreserveBackslash\centering}m{\cellWidtha}>{\PreserveBackslash\centering}m{\cellWidthb}>{\PreserveBackslash\centering}m{\cellWidthc}>{\PreserveBackslash\centering}m{\cellWidthd}}
\toprule

\textbf{Observatory}	& \textbf{PCC}	& \textbf{\emph{p}-Value} & \\
\midrule
Chajnantor (APEX)		& -			& $0.385$ & no~correlation \\
Paranal		& $0.134$			& $0.022$ &  anti-correlation\\
la Silla		& $0.380$			& $0.000$ & anti-correlation\\

\bottomrule
\end{tabularx}}
\end{table}
\unskip

\subsection{The Impact on Astronomical~Seeing}

In observational astronomy, and~as a more technical addition to the definition in Section~\ref{sec:intro}, seeing is a term that refers to the broadening in the point-spread-function (PSF) of a point-like source (a star) attributed to atmospheric optical turbulence. The~broadening occurs due to the scattering of light away from the core of the PSF because of rapid fluctuations in the air index of refraction along the path the starlight follows through the atmosphere to an imaging~detector. 

In optical wavelengths, the index of refraction fluctuation is dominated by the turbulent mixing of air of different temperatures~\cite{Ciddor1996}. The effects of turbulent fluctuations of humidity in the air index of refraction are more relevant for light in the infrared and radio wavelengths~\cite{Thayer1974}. In~this paper, we refer to seeing, in~the optical band, more specifically as measured by the differential image motion monitor (DIMM) instrument at a wavelength of 500~nm. In~simple terms, seeing is a measure of the angular size, in~arcsecs, of~the PSF at the half-intensity point and at the first order driven by changes in wind direction and temperature gradients. 
While it is understood in detail how seeing is influenced by environmental parameters, predictive tools for seeing on longer time scales are not available to the community.

\begin{figure}[H]
\includegraphics[width=13.8cm]{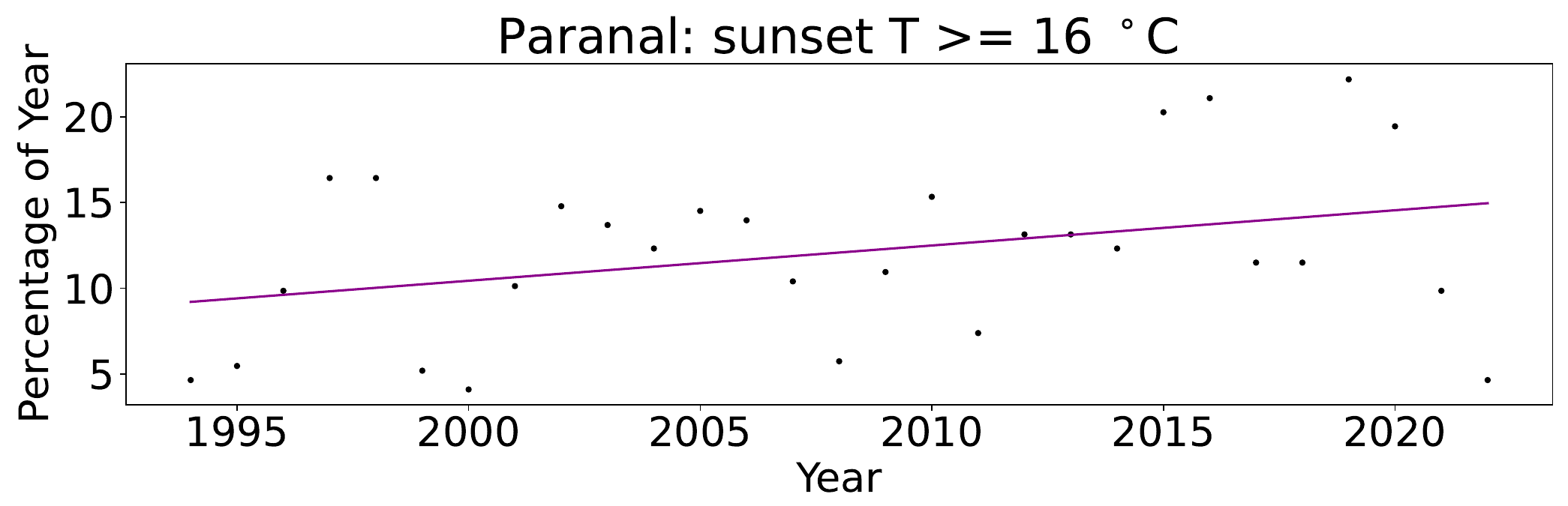}
\caption{The fraction of the year where the sunset air temperature at 2~m is above 16~$^\circ$C as black dots, expanding Figure~\ref{fig:PWVParanal}c from~\cite{Cantalloube2020}, which spanned the years 2008--2020. The~magenta line marks a linear fit to the data to indicate the long-term~trend. \label{fig:ParAbove16}}
\end{figure}

\begin{figure}[H]
\includegraphics[width=13.8cm]{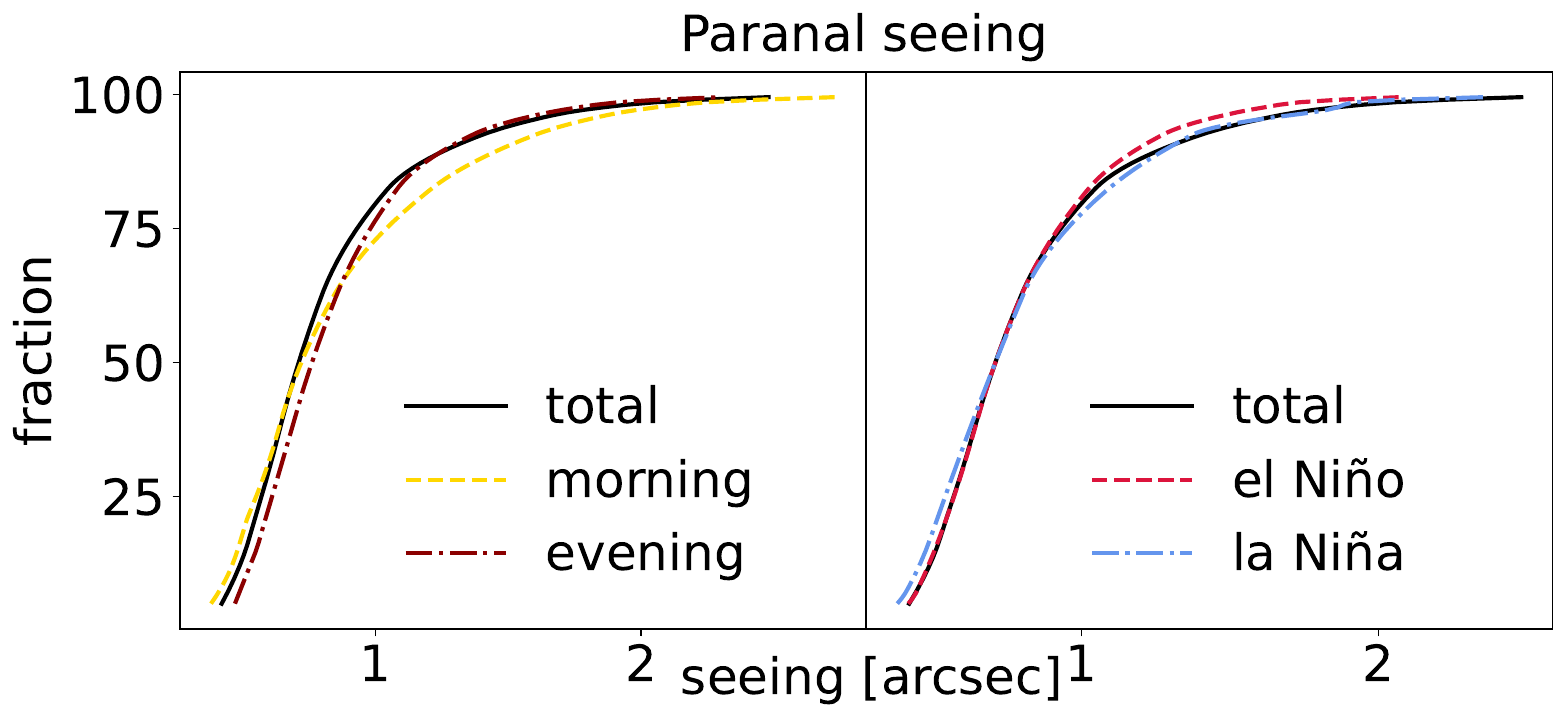}
\caption{Mean seeing probability density function (PDF) from Table~\ref{tab:seeingtwilight} comparing morning and evening seeing on the left and the same normalised function for strong EN and LN events from Tables~\ref{tab:seeingEN} and \ref{tab:seeingLN}. The~PDF, on~the right-hand side, is normalised such that all curves match the long-term median of seeing at the Paranal Observatory. The~figure helps to compare the relative shapes of the probability density function~curves.} \label{fig:fracTwilight}
\end{figure}

In the context of the ELT, the~impact of climate variations on atmospheric turbulence and, thus, seeing at the Paranal Observatory is of the utmost importance. As stated above, the~temperature gradient has an important impact on seeing. This gradient is naturally largest at sunset when the surface and telescope domes are heated from the strong solar irradiation in the Atacama desert. The authors of \citep{Cantalloube2020} studied the amount of time each year that the sunset temperature was above 16~$^\circ$C for the period 2008--2020. The value of 16~$^\circ$C is the maximum temperature at which the telescope dome environments can be maintained while still providing safe cooling to the electronics equipment. We have expanded the analysis to include the full available time series 1994 to 2022 and confirm their trends (see Figure~\ref{fig:ParAbove16}). The~yearly fraction of sunset surface air temperatures exceeding  16~$^\circ$C has increased at a rate of over $2\%$ per decade, reaching up to a 20--25\% yearly fraction during the last EN phase in 2018--2019. This trend is in line with global warming, as confirmed via the MK statistical test. Because~of the necessary cooling within the telescope dome environments, on~warmer days (temperatures at sunset above 16~$^\circ$C), the~domes are cooler than the external ambient temperature at the time of dome opening. In~\citep{Cantalloube2020}, it is inferred that this particular temperature gradient leads to stronger turbulence. However, we may argue that the internal and cooler atmosphere inside the dome is stratified and convection is, thus, inhibited. That said, this may require a specific study on the science image quality as a function of the difference in temperature between the internal dome environment and the external air. At~least, in~terms of atmospheric seeing, our study of evening and morning twilight seeing compared to total seeing for the most affected years (2016--2021) shows that the fraction of high-seeing events ($>$1.6~arcseconds) for the evening twilight is equal to the total seeing statistics (see Figure~\ref{fig:fracTwilight} and Table~\ref{tab:seeingtwilight}). 

\begin{figure}[H]
\includegraphics[width=13.8cm]{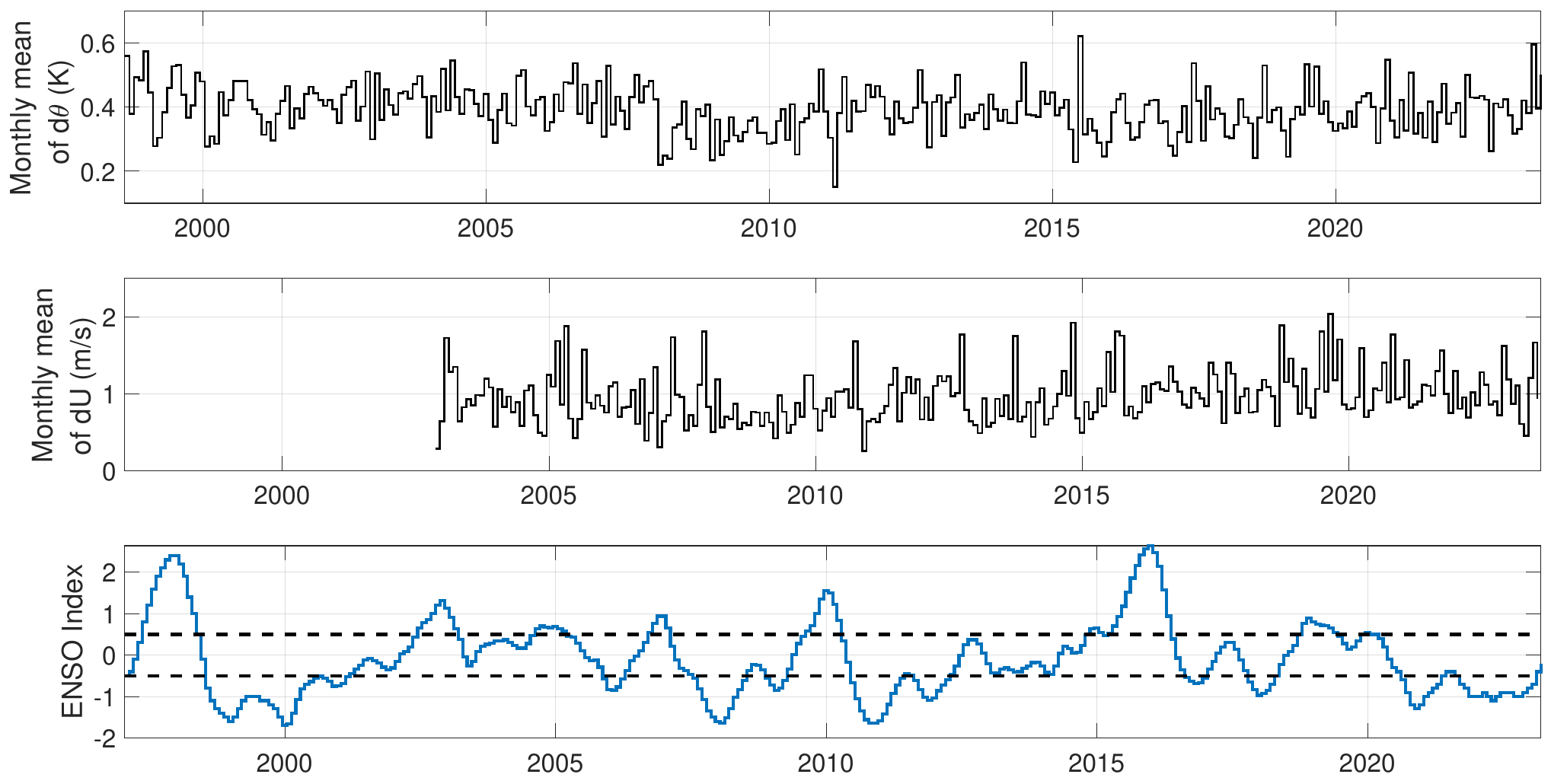}
\caption{(\textbf{top}) Monthly means of the air temperature difference between measurements at 30~m and 2~m above ground level. (\textbf{middle}) Monthly means of the differences in the horizontal wind speed measured at 30~m and 10~m above ground level. The~\textbf{bottom} plots, for completeness, show the time series of the El Niño index. Note: The trends of wind differences are shown starting in 2002. The~wind record prior to 2002 shows some inconsistencies that produced a very small magnitude wind gradient. It is possible that the measurements for the period prior to 2002 were not performed exactly  at the heights above ground as reported in the database, and,~therefore, we opted to exclude this~data.} \label{fig:tempwindgrad}
\end{figure}

Furthermore, and~motivated by the work of~\cite{Cantalloube2020}, we analysed the near surface level temperature record to check for possible trends in temperature gradients between the 30~m and 2~m height from the ground level. Similarly, we also analysed the vertical gradient in the horizontal wind speed from wind measurements performed at 30~m and 10~m above ground level. The~long-term record for the monthly means of the vertical differences in temperature and horizontal wind speed are shown in Figure~\ref{fig:tempwindgrad}. Our results show seasonal variability, but, overall, the temperature and horizontal wind speed gradient have remained stable, within~their natural variability. For~completeness, the~gradients are shown together with the ENSO index. This result on the wind speed and temperature gradient expands the exhaustive work on the relation of seeing and wind speed and temperature gradients in the context of climate change from 2009 at the Paranal Observatory \citep{Melnick2009}. In~that work, the authors analysed the data from 1990 to 1998 and from 1998 to 2008 and concluded that climate change had no significant impact on seeing in this time period. Additionally, and~most importantly, they concluded that the Armazones site, now host of the ELT construction, was unaffected both by seeing changes due to climate change or due to EN or LN phases. We expand their findings with complementary data from the decade following their~study.

\vspace{-5pt}

\begin{figure}[H]
\includegraphics[width=13.8cm]{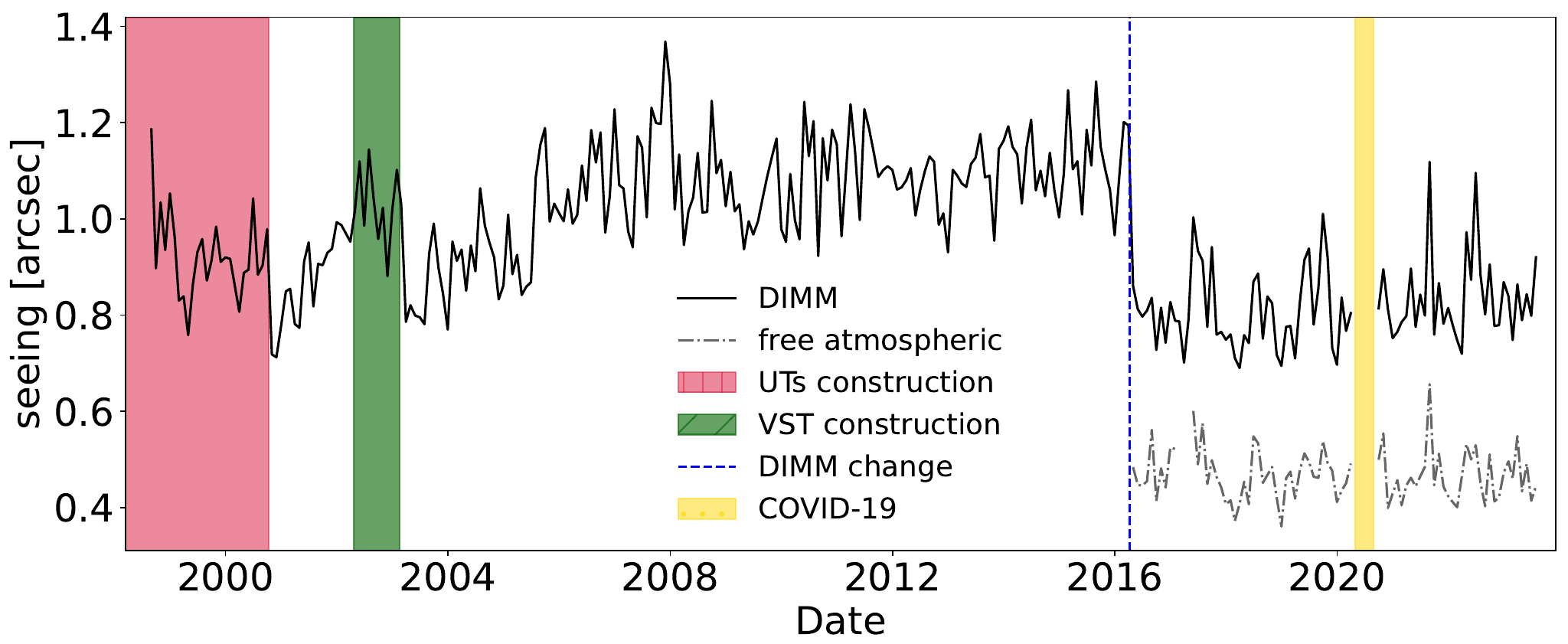}
\caption{Mean monthly seeing at the Paranal Observatory with various important changes highlighted that impacted the mean seeing trend. No net increase in seeing is discernible at this~stage.} \label{fig:seeingParanal}
\end{figure}

{To better understand the potential impact of ENSO on seeing}, we studied the seeing during the night, separating out especially strong EN/LN phases (see \mbox{Appendix~\ref{app:seeing}}-Tables \ref{tab:seeingEN} and \ref{tab:seeingLN}) and found no significant difference in seeing, although~dry EN phases showed less variance in seeing. Figure~\ref{fig:seeingParanal} shows the seeing at the Paranal Observatory in the period 1998-2023. This time series is affected by biases induced by construction and the proximity of the telescope buildings to the former seeing monitor tower (1998-April~2016)~\citep{Cantalloube2020}, a~position prone to the influence of surface-layer turbulence re-circulation. The~new seeing monitor instrument (DIMM) was then moved away from the telescope building and raised up, from~5~m above ground level to 7~m. Due to this improvement, the~seeing came back to the best long-term values before construction, showing that no gradual increase with time was taking place. In~the latest record, we may see evidence of an increase in the lowest seeing values, but~the time series is still too short to attempt a correlation with ENSO or climate change.

Moreover, we explored a possible trend between the surface-level turbulence differences between the EN and LN phases of ENSO via the Richardson number. Our results, derived from the data in Figure~\ref{fig:tempwindgrad} and taking $\theta$ = 314 K in Equation~(\ref{eq:richardson}), show a nearly constant long-term Richardson number of about 0.2--0.3, consistent with the earlier work of~\cite{Melnick2009}, and~no differences between the EN and LN~periods.

As well as the integrated (whole-atmosphere) seeing in Figure~\ref{fig:seeingParanal}, we have included the free-atmosphere seeing from 2016 onwards, which does not show any particular trend either. Free-atmosphere seeing results from optical turbulence from 250~m onwards, away from the surface layer, up~to the top of the troposphere. The~MASS instrument allows measuring the free-atmosphere seeing~\cite{Tokovinin2007}, as~well as other parameters important for observational~astronomy.

The period after 2020 was associated with LN conditions, and~our results, as~shown in Appendix~\ref{app:seeing}, indicate a higher seeing variability during LN periods. As~we are entering (2023) into the next EN period, the~future record of seeing will show if the lowest seeing conditions decrease compared to the 2016-2020 period characterised by mainly neutral ENSO conditions. Nonetheless, it is important to state that the seeing conditions in the Paranal and Cerro Armazones region (ELT site) remain exceptional for seeing-limited astronomical observations and, at~large, are unaffected by ENSO and climate change, confirming the results from a decade prior \citep{Melnick2009}.

\section{Conclusions}
\label{sec:concl}
From the analysis of meteorological data gathered at the European Southern Observatory sites, we found a positive correlation of extremely surface-level warm periods with ENSO across all sites, as~well as an anti-correlation between ENSO and the surface-level absolute humidity. In~other words, during~EN phases, there was a high likelihood of warmer and dryer surface conditions. Similarly, during~LN, the~surface meteorological conditions call for lower maximum daily temperatures but higher absolute humidity. As a consequence, we now, for~the first time, have predicting capabilities to better schedule more demanding infrared observation programs that cannot be executed in elevated humidity. Conversely, during~LN phases, when periods of wet and cloudy conditions are more probable, we can plan for less restrictive programs.

Unrelated to astronomy, we confirm the overall trend of increasing temperatures for all sites with rates of about or more than $0.2^\circ$C/decade which are consistent with anthropogenic climate change since 1980. Our results show a significant increase in temperatures at the Chajnantor site in line with the acceleration of warming with altitude. These long-term, high-altitude stations built for astronomy could potentially become an important resource for the climate community as they fill an important gap in station coverage ~\cite{Mutz2021}. Regarding precipitable water vapour, our data for the Paranal Observatory do not show an increase in any metrics with time, confirming studies that have been conducted at the Chajnantor site. 

While the trend in the absolute surface air temperature is evident, our results do not show a long-term trend in the vertical gradient of air temperature close to the surface, nor in the vertical gradient of the horizontal component of wind speed. Their ratios, following the formalism of the Richardson number, do not show a long trend or significant differences between strong phases of ENSO either, implying that the ground-level turbulence has remained within its natural seasonal variability, in~line with previously conducted studies on seeing. 

Regarding a possible trend in astronomical seeing since then, unfortunately, our long-term dataset for the Paranal Observatory is affected by biases that are hard to correct without introducing systematic errors. Nonetheless, we find no evidence for an increase in seeing or turbulence with time, contrary to popular beliefs on astronomical data quality over time. However, it seems that the spread in seeing is larger during strong LN phases, making conditions during LN even more unpredictable. The record of seeing in the following years will help elucidate a more direct impact of strong EN and LN events on seeing and we will discuss any visible trends in follow-up work. Additionally, the~impact of the dome seeing on astronomical observations is not currently well studied and should be measured systematically in the future.

In conclusion, our work serves to confirm the increasing warming trend with altitude due to global warming, underscoring the unique exposure of observatory sites to the challenges that lie ahead. Despite not finding any direct impact on seeing yet, our research contributes valuable insights for observatories in their preparation for the challenging future climatic conditions, as well as a predictive tool via the ENSO phases for the scheduling of challenging near-infrared programmes.

\vspace{6pt} 



\authorcontributions{J.V.S. has led the data collection, analysis, and interpretation of the temperature and WVD data, including the specific Paranal analysis of the 16 Celsius Degrees threshold, as well as the overall interpretation of the results. J.V.S. has written the manuscript and is responsible for its contents. A.O. proposed the project, led the discussion for data analysis and interpretation, and produced the WVD data record, contributing also to the writing and proofreading. V.T. led the analysis of the astronomical seeing data at Paranal.
}

\funding{This research has received funding from the European Organisation for Astronomical Research in the Southern Hemisphere (ESO).}

\institutionalreview{Not applicable.}

\informedconsent{Not applicable.}

\dataavailability{\textls[-10]{Climate indices are available at \url{www.psl.noaa.gov/data/climateindices/}}, recent APEX, Paranal, and~la Silla ambient conditions data from: \url{http://archive.eso.org/cms/eso-data/ambient-conditions.html}, CBI data from \url{www.sites.astro.caltech.edu/~tjp/CBI/weather/index.html}. The~PWV Paranal LHATPRO data can be queried at \url{www.archive.eso.org/wdb/wdb/asm/lhatpro_paranal/form}. The~historic PWV Paranal UVES and XSHOOTER data can be retrieved at \url{www.eso.org/observing/dfo/quality/GENERAL/PWV/HEALTH/trend_report_ambient_PWV_closeup_HC.html}. All other historical data are available upon request either from the corresponding author or from the European Southern Observatory climate officer, A. Otarola. All websites were accessed in August 2023.} 

\acknowledgments{The authors thank Alain Smette for retrieving the historic Paranal PVW data and Aldo Pizarro for his knowledge of la Silla's historic events. The~historical PVW data was calculated by Reinhard Hanuschick, the~retired head of ESO's Quality Control, whom we thank for making this data publicly available. This work relied on observations collected at the European Southern Observatory (ESO). We thank the three referees and the editor for their insightful comments which have improved the manuscript.}

\conflictsofinterest{The authors declare no conflict of~interest.} 


\abbreviations{Abbreviations}{
The following abbreviations are used in this manuscript:\\

\noindent 
\begin{tabular}{@{}ll}
AAO & Antarctic circulation\\
AMBL & atmospheric marine boundary layer\\
APEX & Atacama Pathfinder Experiment\\
CBI & Cosmic Background Imager\\
\end{tabular}

\noindent 
\begin{tabular}{@{}ll}
ESO & European Southern Observatory\\
ENSO & El Niño Southern Oscillation\\
FT & free troposphere\\
DJF & season December, January, February (austral summer)\\
JJA & season June, July, August (austral winter)\\
MAM & season March, April, May  (austral autumn)\\
MASL & metres above sea level\\
NEO & near-Earth object\\
PCC & Pearson's correlation coefficient\\
PDO & Pacific decadal oscillation\\
PWV & precipitable water vapour\\
SON & season September, October, November (austral spring)\\
SST & sea surface temperature\\
TI & temperature inversion\\
WVD & water vapour density
\end{tabular}
}

\appendixtitles{yes} 
\appendixstart
\appendix
\section[\appendixname~\thesection]{MK-Test of Temperature Trends across European Southern Observatory~Sites}\label{app:MKtest}
Given our limitation in la Silla Observatory data to 2010 onwards (see Table~\ref{tab:overviewData}), we have split the Paranal Observatory data into pre- and post-2010. The~pre-2010 Paranal Observatory temperature data, spanning approximately 15 years from 1994 to 2009, exhibit decreasing temperature trends (compare Tables~\ref{tab:ParanalT} and \ref{tab:ParanalTcurrent}). The~hiatus in warming during the 1990s and the first decade of the 21st century for Chilean coastal and valley sites below 800 MASL is a well-studied phenomenon \citep{Falvey2009, Burger2018}, most notably for the maximum temperature as observed during autumn in the Paranal Observatory with a strong negative slope. This cooling is related  to a shift to the negative phase of the Interdecadal Pacific Oscillation (IPO), caused by global warming \citep{Falvey2009}, which has ended since then. This trend has only been observed at altitudes below the Paranal Observatory; however, the Paranal Observatory is less than 20 km direct distance from the coast. The~inversion layer that separates the coastal climate from the higher altitudes of the Paranal Observatory is strongest in JJA, where we see no decreasing trends for the pre-2010 Paranal Observatory data. We hypothesize that the coastal cooling trend has only influenced the Paranal Observatory during periods with a weaker atmospheric~stratification.  We have highlighted the most prominent temperature trends with boldface in the tables.
\begin{table}[H]
\caption{Temperature trends Llano del Chajnantor site, 2006--2023.\label{tab:APEXT}}
\setlength{\cellWidtha}{\columnwidth/6-2\tabcolsep-0.0in}
\setlength{\cellWidthb}{\columnwidth/6-2\tabcolsep-0.0in}
\setlength{\cellWidthc}{\columnwidth/6-2\tabcolsep-0.0in}
\setlength{\cellWidthd}{\columnwidth/6-2\tabcolsep-0.0in}
\setlength{\cellWidthe}{\columnwidth/6-2\tabcolsep-0.0in}
\setlength{\cellWidthf}{\columnwidth/6-2\tabcolsep-0.0in}
\scalebox{1}[1]{\begin{tabularx}{\columnwidth}{>{\PreserveBackslash\centering}m{\cellWidtha}>{\PreserveBackslash\centering}m{\cellWidthb}>{\PreserveBackslash\centering}m{\cellWidthc}>{\PreserveBackslash\centering}m{\cellWidthd}>{\PreserveBackslash\centering}m{\cellWidthe}>{\PreserveBackslash\centering}m{\cellWidthf}}
\toprule

			\textbf{Season}	& 	& \textbf{Trend}	& \textbf{\emph{p}-Value}  & \textbf{$\tau$}  & \textbf{Slope}   \\
			\midrule
\multirow{3}{*}{\textbf{DJF (day)}}	& \textbf{mean}			& increasing			& $0.001$        &   $0.221$  & $0.036$\\
			  	                   & \textbf{max}			& increasing			& $0.000$     & $0.257$   & $0.055$\\
			             	      & \textbf{min}			& increasing			& $0.003$     & $0.235$    & $0.045$\\
\multirow{3}{*}{\textbf{DJF (night)}}	& \textbf{mean}			& increasing			& $0.000$     & $0.250$    & $0.040$\\
			  	                   & \textbf{max}			& increasing			& $0.000$     & $0.353$    & $0.069$\\
			             	      & \textbf{min}			& increasing			& $0.025$     & $0.176$   & $0.043$\\
                   \midrule
\multirow{3}{*}{\textbf{MAM (day)}}	& mean			& no trend			& $0.278$     & -    & -\\
			  	                   & max			& no trend			& $0.888$     & -   & -\\
			             	      & \textbf{min}			& increasing			& $0.000$     & $0.250$    & $0.094$\\
\multirow{3}{*}{\textbf{MAM (night)}}	& mean			& no trend			& $0.412$     & -    & -\\
			  	                   & max			& no trend			& $1.000$     & -    & -\\
			             	      & \textbf{min}			& increasing			& $0.000$     & $0.221$    & $0.112$\\
			             	      			\bottomrule
		\end{tabularx}}
\end{table}

\begin{table}[H]\ContinuedFloat
\caption{{\em Cont.} \label{tab:APEXT}}
\setlength{\cellWidtha}{\columnwidth/6-2\tabcolsep-0.0in}
\setlength{\cellWidthb}{\columnwidth/6-2\tabcolsep-0.0in}
\setlength{\cellWidthc}{\columnwidth/6-2\tabcolsep-0.0in}
\setlength{\cellWidthd}{\columnwidth/6-2\tabcolsep-0.0in}
\setlength{\cellWidthe}{\columnwidth/6-2\tabcolsep-0.0in}
\setlength{\cellWidthf}{\columnwidth/6-2\tabcolsep-0.0in}
\scalebox{1}[1]{\begin{tabularx}{\columnwidth}{>{\PreserveBackslash\centering}m{\cellWidtha}>{\PreserveBackslash\centering}m{\cellWidthb}>{\PreserveBackslash\centering}m{\cellWidthc}>{\PreserveBackslash\centering}m{\cellWidthd}>{\PreserveBackslash\centering}m{\cellWidthe}>{\PreserveBackslash\centering}m{\cellWidthf}}
\toprule

			\textbf{Season}	& 	& \textbf{Trend}	& \textbf{\emph{p}-Value}  & \textbf{$\tau$}  & \textbf{Slope}   \\
                    \midrule
\multirow{3}{*}{\textbf{JJA (day)}}	& mean			& no trend			& $0.321$     & -    & -\\
			  	                   & max			& no trend			& $0.778$     & -    & -\\
			             	      & \textbf{min}			& increasing			& $0.006$     & $0.213$    & $0.027$\\
\multirow{3}{*}{\textbf{JJA (night)}}	& \textbf{mean}			& increasing			& $0.027$     & $0.176$    & $0.029$\\
			  	                   & max			& no trend			& $0.092$     & -    & -\\
			             	      & min			& no trend			& $0.067$     & -    & -\\

                   \midrule
\multirow{3}{*}{\textbf{SON (day)}}	& \textbf{mean}			& increasing			& $0.045$     &   $0.150$  & $0.026$\\
			  	                   & max			& no trend			& $0.292$     & -   & -\\
			             	      & min			& no trend			& $0.079$     & -    & -\\
\multirow{3}{*}{\textbf{SON (night)}}	& \textbf{mean}			& increasing			& $0.001$     & $0.267$    & $0.049$\\
			  	                   & \textbf{max}			& increasing			& $0.009$     & $0.200$   & $0.025$\\
			             	      & \textbf{min}			& increasing			& $0.012$     & $0.200$   & $0.050$\\
			\bottomrule
		\end{tabularx}}
\end{table}
\unskip

\begin{table}[H]
\caption{Temperature trends Paranal Observatory, 1994--2009.\label{tab:ParanalT}}
\setlength{\cellWidtha}{\columnwidth/6-2\tabcolsep-0.0in}
\setlength{\cellWidthb}{\columnwidth/6-2\tabcolsep-0.0in}
\setlength{\cellWidthc}{\columnwidth/6-2\tabcolsep-0.0in}
\setlength{\cellWidthd}{\columnwidth/6-2\tabcolsep-0.0in}
\setlength{\cellWidthe}{\columnwidth/6-2\tabcolsep-0.0in}
\setlength{\cellWidthf}{\columnwidth/6-2\tabcolsep-0.0in}
\scalebox{1}[1]{\begin{tabularx}{\columnwidth}{>{\PreserveBackslash\centering}m{\cellWidtha}>{\PreserveBackslash\centering}m{\cellWidthb}>{\PreserveBackslash\centering}m{\cellWidthc}>{\PreserveBackslash\centering}m{\cellWidthd}>{\PreserveBackslash\centering}m{\cellWidthe}>{\PreserveBackslash\centering}m{\cellWidthf}}
\toprule
			\textbf{Season}	& 	& \textbf{Trend}	& \textbf{\emph{p}-Value}  & \textbf{$\tau$}  & \textbf{Slope}   \\
			\midrule
\multirow{3}{*}{\textbf{DJF (day)}}	& mean			& no trend			& $0.111$        &   -  &   - \\
			  	                   & max			& increasing			& $0.004$     & $0.361$   & $0.110$\\
			             	      & \textbf{min}			& decreasing			& $0.013$     & $-0.190$   & $-0.030$\\
\multirow{3}{*}{\textbf{DJF (night)}}	& mean			& no trend			& $0.515$     & -    & - \\
			  	                   & max			& increasing			& $0.012$     & $0.257$    & $0.091$ \\
			             	      & \textbf{min}			& decreasing			& $0.000$     & $-0.324$    & $-0.083$\\
                   \midrule
\multirow{3}{*}{\textbf{MAM (day)}}	& mean			& no trend & $0.135$     & - & - \\
			  	                   & \textbf{max}			& decreasing			& $0.000$     & $-0.253$   & $-0.095$\\
			             	      & min			& increasing			& $0.003$     & $0.198$    & $0.078$\\
\multirow{3}{*}{\textbf{MAM (night)}}	& mean			& no trend			& $0.676$     & -    & -\\
			  	                   & \textbf{max}			& decreasing			& $0.029$     & $-0.187$    & $-0.070$\\
			             	      & min			& no trend			& $0.222$     & -    & - \\
                    \midrule
\multirow{3}{*}{JJA (day)}	& mean			& no trend			& $0.496$     & -    & -\\
			  	                   & max			& 		no trend	& $0.162$     & -    & - \\
			             	      & min			& no trend			& $0.205$     & -    & -\\
\multirow{3}{*}{JJA (night)}	& mean			& no trend			& $0.062$     & -    & - \\
			  	                   & max			& no trend			& $0.122$     & -    & - \\
			             	      & min			& increasing			& $0.010$     & $0.282$   & $0.196$ \\
                   \midrule
\multirow{3}{*}{SON (day)}	& mean			& no trend			& $0.205$     &   -  & - \\
			  	                   & max			& no trend			& $0.565$     & -   & - \\
			             	      & min			& no trend			& $0.403$     & -    & - \\
\multirow{3}{*}{SON (night)}	& mean			& no trend			& $0.614$     & -    & - \\
			  	                   & max			& no trend			& $0.428$     & -    & -  \\
			             	      & min			& no trend			& $0.168$     & -    & - \\
			\bottomrule
		\end{tabularx}}
\end{table}

\vspace{-9pt}
\begin{table}[H]
\caption{Temperature trends Paranal Observatory, 2010 -~2023.\label{tab:ParanalTcurrent}}
		\newcolumntype{C}{>{\centering\arraybackslash}X}
\setlength{\cellWidtha}{\columnwidth/6-2\tabcolsep-0.0in}
\setlength{\cellWidthb}{\columnwidth/6-2\tabcolsep-0.0in}
\setlength{\cellWidthc}{\columnwidth/6-2\tabcolsep-0.0in}
\setlength{\cellWidthd}{\columnwidth/6-2\tabcolsep-0.0in}
\setlength{\cellWidthe}{\columnwidth/6-2\tabcolsep-0.0in}
\setlength{\cellWidthf}{\columnwidth/6-2\tabcolsep-0.0in}
\scalebox{1}[1]{\begin{tabularx}{\columnwidth}{>{\PreserveBackslash\centering}m{\cellWidtha}>{\PreserveBackslash\centering}m{\cellWidthb}>{\PreserveBackslash\centering}m{\cellWidthc}>{\PreserveBackslash\centering}m{\cellWidthd}>{\PreserveBackslash\centering}m{\cellWidthe}>{\PreserveBackslash\centering}m{\cellWidthf}}
\toprule
			\textbf{Season}	& 	& \textbf{Trend}	& \textbf{\emph{p}-Value}  & \textbf{$\tau$}  & \textbf{Slope}   \\
			\midrule
\multirow{3}{*}{\textbf{DJF (day)}}	& mean			& no trend			& $0.111$        &   -  &   - \\
			  	                   & max			& increasing			& $0.004$     & $0.362$   & $0.110$\\
			             	      & \textbf{min}			& no trend			& $0.261$     & -    & -\\
\multirow{3}{*}{\textbf{DJF (night)}}	& mean			& no trend			& $0.515$     & -    & - \\
			  	                   & max			& increasing			& $0.012$     & $0.257$   & $0.091$ \\
			             	      & \textbf{min}			& no trend			& $0.102$     & -    & -\\
                   \midrule
\multirow{3}{*}{\textbf{MAM (day)}}	& mean			& no trend & $0.135$     & - & - \\
			  	                   & max			& decreasing			& $0.001$     & $-0.187$   & $-0.095$\\
			             	      & \textbf{min}			& increasing			& $0.003$     & $0.231$    & $0.078$\\
			             	          \bottomrule
		\end{tabularx}}
\end{table}

\begin{table}[H]\ContinuedFloat
\caption{{\em Cont.}\label{tab:ParanalTcurrent}}
		\newcolumntype{C}{>{\centering\arraybackslash}X}
\setlength{\cellWidtha}{\columnwidth/6-2\tabcolsep-0.0in}
\setlength{\cellWidthb}{\columnwidth/6-2\tabcolsep-0.0in}
\setlength{\cellWidthc}{\columnwidth/6-2\tabcolsep-0.0in}
\setlength{\cellWidthd}{\columnwidth/6-2\tabcolsep-0.0in}
\setlength{\cellWidthe}{\columnwidth/6-2\tabcolsep-0.0in}
\setlength{\cellWidthf}{\columnwidth/6-2\tabcolsep-0.0in}
\scalebox{1}[1]{\begin{tabularx}{\columnwidth}{>{\PreserveBackslash\centering}m{\cellWidtha}>{\PreserveBackslash\centering}m{\cellWidthb}>{\PreserveBackslash\centering}m{\cellWidthc}>{\PreserveBackslash\centering}m{\cellWidthd}>{\PreserveBackslash\centering}m{\cellWidthe}>{\PreserveBackslash\centering}m{\cellWidthf}}
\toprule
			\textbf{Season}	& 	& \textbf{Trend}	& \textbf{\emph{p}-Value}  & \textbf{$\tau$}  & \textbf{Slope}   \\
\multirow{3}{*}{\textbf{MAM (night)}}	& mean			& no trend			& $0.676$     & -    & -\\
			  	                   & max			& decreasing			& $0.029$     & $-0.187$    & $-0.054$\\
			             	      & \textbf{min}			& increasing			& $0.002$     & $0.253$    & $0.230$\\

                    \midrule
\multirow{3}{*}{JJA (day)}	& mean			& no trend			& $0.917$     & -    & -\\
			  	                   & max			& 		no trend	& $0.162$     & -    & - \\
			             	      & min			& no trend			& $0.205$     & -    & -\\
			\midrule
\multirow{3}{*}{JJA (night)}	& mean			& no trend			& $0.918$     & -    & - \\
			  	                   & max			& no trend			& $0.702$     & -    & - \\
			             	      & min			& increasing			& $0.010$     & $0.282$   & $0.196$ \\
                   \midrule
\multirow{3}{*}{SON (day)}	& mean			& no trend			& $0.609$     &   -  & - \\
			  	                   & max			& no trend			& $0.410$     & -   & - \\
			             	      & min			& no trend			& $0.089$     & -    & -\\
\multirow{3}{*}{SON (night)}	& mean			& no trend			& $0.714$     & -    & - \\
			  	                   & max			& no trend			& $0.168$     & -    & -  \\
			             	      & min			& no trend			& $0.212$     & -    & -\\
			\bottomrule
		\end{tabularx}}
\end{table}
\unskip

\begin{table}[H]
\caption{Temperature trends la Silla Observatory, 2010--2023.\label{tab:laSillaT}}
		\newcolumntype{C}{>{\centering\arraybackslash}X}
\setlength{\cellWidtha}{\columnwidth/6-2\tabcolsep-0.0in}
\setlength{\cellWidthb}{\columnwidth/6-2\tabcolsep-0.0in}
\setlength{\cellWidthc}{\columnwidth/6-2\tabcolsep-0.0in}
\setlength{\cellWidthd}{\columnwidth/6-2\tabcolsep-0.0in}
\setlength{\cellWidthe}{\columnwidth/6-2\tabcolsep-0.0in}
\setlength{\cellWidthf}{\columnwidth/6-2\tabcolsep-0.0in}
\scalebox{1}[1]{\begin{tabularx}{\columnwidth}{>{\PreserveBackslash\centering}m{\cellWidtha}>{\PreserveBackslash\centering}m{\cellWidthb}>{\PreserveBackslash\centering}m{\cellWidthc}>{\PreserveBackslash\centering}m{\cellWidthd}>{\PreserveBackslash\centering}m{\cellWidthe}>{\PreserveBackslash\centering}m{\cellWidthf}}
\toprule
			\textbf{Season}	& 	& \textbf{Trend}	& \textbf{\emph{p}-Value}  & \textbf{$\tau$}  & \textbf{Slope}   \\
			\midrule
\multirow{3}{*}{DJF (day)}	& mean			& no trend			& $0.293$        &   -  & -\\
			  	                   & max			& no trend			& $0.084$     & -    & -\\
			             	      & min			& no trend			& $0.421$     & -    & -\\
\multirow{3}{*}{DJF (night)}	& mean			& no trend			& $0.064$     & -    & - \\
			  	                   & max			& no trend			& $0.078$     & -    & -\\
			             	      & min			& no trend			& $0.764$     & -    & -\\
                   \midrule
\multirow{3}{*}{\textbf{MAM (day)}}	& \textbf{mean}			& increasing			& $0.000$     & $0.275$    & $0.168$\\
			  	                   & max			& no trend			& $0.131$     & -   & -\\
			             	      & \textbf{min}			& increasing			& $0.006$     & $0.275$    & $0.267$\\
\multirow{3}{*}{\textbf{MAM (night)}}	& \textbf{mean}			& increasing			& $0.003$     & $0.231$    & $0.153$\\
			  	                   & max			& no trend			& $0.438$     & -    & -\\
			             	      & \textbf{min}			& increasing			& $0.040$     & $0.231$    & $0.290$\\
                    \midrule
\multirow{3}{*}{JJA (day)}	& mean			& no trend			& $0.285$     & -    & - \\
			  	                   & max			& no trend			& $0.746$     & -    & -\\
			             	      & min			& no trend			& $0.380$     & -    & -\\
\multirow{3}{*}{JJA (night)}	& mean			& no trend			& $0.138$     & -    & - \\
			  	                   & max			& no trend			& $0.162$     & -    & -\\
			             	      & min			& no trend			& $0.119$     & -   & -\\
                   \midrule
\multirow{3}{*}{\textbf{SON (day)}}	& \textbf{mean}			& increasing			& $0.039$     &   $0.205$  & $0.061$\\
			  	                   & max			& no trend			& $0.464$     & -   & -\\
			             	      & \textbf{min}			& increasing			& $0.000$     & $0.397$   & $0.382$\\
\multirow{3}{*}{\textbf{SON (night)}}	& mean			& no trend			& $0.325$     & -    & - \\
			  	                   & max			& no trend			& $0.488$     & -    & -\\
			             	      & \textbf{min}			& increasing			& $0.000$     & $0.449$    & $0.256$\\
			\bottomrule
		\end{tabularx}}
\end{table}
\unskip

\section[\appendixname~\thesection]{Paranal Seeing Analysis}
\label{app:seeing}\vspace{-6pt}

\begin{table}[H] 
\caption{Probability density function of mean seeing during the period 2016 to~2021. } \label{tab:seeingtwilight}
\setlength{\cellWidtha}{\columnwidth/4-2\tabcolsep-0.0in}
\setlength{\cellWidthb}{\columnwidth/4-2\tabcolsep-0.0in}
\setlength{\cellWidthc}{\columnwidth/4-2\tabcolsep-0.0in}
\setlength{\cellWidthd}{\columnwidth/4-2\tabcolsep-0.0in}
\scalebox{1}[1]{\begin{tabularx}{\columnwidth}{>{\PreserveBackslash\centering}m{\cellWidtha}>{\PreserveBackslash\centering}m{\cellWidthb}>{\PreserveBackslash\centering}m{\cellWidthc}>{\PreserveBackslash\centering}m{\cellWidthd}}
\toprule

\textbf{Fraction}	& \textbf{Total Nights}	& \textbf{Morning Twilight} & \textbf{Evening Twilight}\\
\midrule
$5\%$		& $0.42$			& $0.38$ & $0.47$ \\
$10\%$		& $0.47$			& $0.44$ &  $0.51$\\
$15\%$		& $0.51$			& $0.48$ & $0.55$\\
$20\%$		& $0.54$			& $0.51$ & $0.58$\\
$25\%$		& $0.57$			& $0.55$ & $0.61$ \\
$50\%$		& $0.71$			& $0.72$ &  $0.76$\\
$75\%$		& $0.93$			& $1.04$ & $0.98$\\
$90\%$		& $1.28$			& $1.48$ & $1.27$\\
$95\%$		& $1.57$			& $1.77$ & $1.51$\\
$97.5\%$	& $1.84$			& $2.04$ & $1.75$\\
$99\%$		& $2.22$			& $2.43$ & $2.06$\\
$99.5\%$	& $2.48$			& $2.73$ & $2.3$\\
\bottomrule
\end{tabularx}}
\end{table}
\unskip

\begin{table}[H] 
\caption{Spread of seeing for strong EN phases ($\delta T > 0.75\,^\circ\text{C}$).} \label{tab:seeingEN}
\setlength{\cellWidtha}{\columnwidth/4-2\tabcolsep+0.6in}
\setlength{\cellWidthb}{\columnwidth/4-2\tabcolsep-0.2in}
\setlength{\cellWidthc}{\columnwidth/4-2\tabcolsep-0.2in}
\setlength{\cellWidthd}{\columnwidth/4-2\tabcolsep-0.2in}
\scalebox{1}[1]{\begin{tabularx}{\columnwidth}{>{\PreserveBackslash\centering}m{\cellWidtha}>{\PreserveBackslash\centering}m{\cellWidthb}>{\PreserveBackslash\centering}m{\cellWidthc}>{\PreserveBackslash\centering}m{\cellWidthd}}
\toprule
\textbf{Period}	& \textbf{Variance}	& \textbf{75--25\%} & \textbf{95--5\%}\\
\midrule
2-July to December		& $0.13$			& $0.42$ & $1.14$ \\
6-October to November		& $0.28$			& $0.55$ &  $1.64$\\
9-October to 10-February		& $0.20$			& $0.47$ & $1.37$\\
15-May to 16-March		& $0.19$			& $0.50$ & $1.36$\\
\bottomrule
\end{tabularx}}
\end{table}
\unskip

\begin{table}[H] 
\caption{Spread of seeing for strong LN phases ($\delta T < -0.75\,^\circ\text{C}$).} \label{tab:seeingLN}
\setlength{\cellWidtha}{\columnwidth/4-2\tabcolsep+0.6in}
\setlength{\cellWidthb}{\columnwidth/4-2\tabcolsep-0.2in}
\setlength{\cellWidthc}{\columnwidth/4-2\tabcolsep-0.2in}
\setlength{\cellWidthd}{\columnwidth/4-2\tabcolsep-0.2in}
\scalebox{1}[1]{\begin{tabularx}{\columnwidth}{>{\PreserveBackslash\centering}m{\cellWidtha}>{\PreserveBackslash\centering}m{\cellWidthb}>{\PreserveBackslash\centering}m{\cellWidthc}>{\PreserveBackslash\centering}m{\cellWidthd}}
\toprule
\textbf{Period}	& \textbf{Variance}	& \textbf{75--25\%} & \textbf{95--5\%}\\
\midrule
98-July to 00-March		& $0.19$			& $0.45$ & $1.39$ \\
5-December to 6-January		& $0.13$			& $0.40$ &  $1.12$\\
7-August to 8-April		& $0.24$			& $0.59$ & $1.46$\\
8-December to 9-January		& $0.24$			& $0.55$ & $1.46$\\
10-July to 11-February		& $0.34$			& $0.54$ & $1.66$ \\
11-September to 11-December		& $0.18$			& $0.48$ &  $1.27$\\
17-November to 18-January		& $0.12$			& $0.33$ & $1.10$\\
20-September to 21-February		& $0.13$			& $0.38$ & $1.12$\\
\bottomrule
\end{tabularx}}
\end{table}
\begin{adjustwidth}{-\extralength}{0cm}

\reftitle{References}

\PublishersNote{}
\end{adjustwidth}
\end{document}